\documentclass[aps,prb,superbib,superscriptaddress,amsfonts,preprint]{revtex4}

\usepackage{epsfig,amsmath,amssymb,graphicx,amscd, soul,units, float}
\usepackage[papersize={8.5in,11in}]{geometry}
\usepackage{xcolor}
\usepackage{hyperref}
\hypersetup{
    colorlinks = true,
    linkcolor = blue,
    linkbordercolor = white,
    citecolor=blue,
}

\newcommand{\ket}[1]{\left| #1 \right>}

\newcommand{\avg}[1]{\ensuremath{\left\langle #1 \right\rangle}}

\newcommand{\vR}{{\vec{R}}}

\begin{document}

\author{Joseph Subotnik\footnote{subotnik@sas.upenn.edu}}
\author{Gaohan Miao}
\author{Nicole Bellonzi}
\author{Hung-Hsuan Teh}
\affiliation{Dept. of Chemistry, University of Pennsylvania}
\author{Wenjie Dou}
\affiliation{Dept. of Chemistry, University of California, Berkeley}

\begin{abstract}
Although the quantum classical Liouville equation (QCLE) arises by cutting off the exact equation
of motion for a coupled nuclear-electronic system at order 1 $(1 = \hbar^0$), we show that the QCLE does include Berry's phase effects and Berry's forces (which are proportional to a higher order, $\hbar = \hbar^1$). Thus, the fundamental equation underlying mixed quantum-classical dynamics does not need a correction
for Berry's phase effects and is valid for the case of complex Hamiltonians. Furthermore, we also show that, even though Tully's surface hopping  model ignores Berry's phase, Berry's phase effects are included
 automatically within Ehrenfest dynamics. These findings should be of great importance if we seek to model coupled nuclear-electronic dynamics for systems with spin-orbit coupling, where the complex nature of the Hamiltonian is paramount.
\end{abstract}

\title{A Demonstration of Consistency between the Quantum Classical Liouville Equation and Berry's Phase and Curvature}
\maketitle

\section{Introduction}
Nonadiabatic dynamics are a continuous source of interest and intrigue in the chemical physics community. On the one hand, 
the fast exchange of energy between nuclear and electronic degrees of freedom violates the Born Oppenheimer (BO) approximation, the bedrock of modern chemistry\cite{szabo:ostlund}.  When one violates the BO approximation even moderately, one can 
find many unexpected effects, the most famous being Berry's phase effects\cite{berry:1984:berryphase}.
On the other hand, because quantum mechanics is so expensive to propagate, there is a strong impetus to understand
nonadiabatic dynamics in a semiclassical fashion\cite{tully:fssh,doltsinis:2002:review,subotnik:2016:arpc,tretiak:2014:acr,martinez:1996:jpc}, focusing on quantum electrons and classical nuclei. Thus, for many researchers,
the nature of nonadiabatic effects becomes entangled with semiclassical approximations, which leads to only more questions 
about the fundamental nature of nonadiabatic dynamics.


In the present communication, we want to directly address one such fundamental question in nonadiabatic dynamics: 
the connection between Berry's phase\cite{berry:1984:berryphase} 
and the quantum classical Liouville equation\cite{kapral:1999:jcp,martens:1997:partwig}. A few words are now appropriate
regarding Berry's phase, both in the context of real and complex Hamiltonians.
In general, Berry's phase effects are usually derived by considering the phase of an electronic wavefunction
in the limit of a very slowly evolving potential that mixes together different adiabatic states,
and the presence of Berry's phase can lead to interference effects around degeneracies(e.g. the Aharanov-Bohm effect\cite{aharonovbohm:1959,yarkony:berry:ch2_dipole} 
and tunneling suppression\cite{guo_yarkony:2016:jacs}).
When the Hamiltonian is real,  Berry's phase is effectively a generalization of the Longuet-Higgins phase\cite{longuet:1958:phase,longuet:1961:phase,longuet:1963:phase},
and there is an enormous literature in the chemical physics literature regarding the role
of Berry's phase effects around conical intersections \cite{yarokny:1996:rmp,ryabinkin:2017:ci_berryphase}.  Of note, however, is that for a complex Hamiltonian,
Berry's phase can yield real effects even without a relevant intersection point; the Berry curvature 
(see equation \ref{eq_berry_force}) will be nonzero\cite{meadtruhlar:1979:berry,shankarbook}.
Although this case is not usually addressed in the chemical physics literature (where we usually assume
that the molecular Hamiltonian is real), the question of curve crossings with complex Hamiltonians has been investigated previously.\cite{
mead:1979:noncrossing,
matsika_yarkony1:2001:jcp,
matsika_yarkony2:2001:jcp,
matsika_yarkony3:2002:jcp} and Takatsuka and Yonehara have written extensively
about Berry's ``Lorentz-like'' forces in the context of semiclassical, path branching dynamics
\cite{takatsuka:2011:pccp_review_nonadiabatic,takatsuka:2010:acp_nonadiabatic}.

Let us now turn to the QCLE\cite{kapral:1999:jcp,martens:1997:partwig}.  The QCLE represents the simplest means to rigorously
take the semiclassical limit of a coupled nuclear-electronic systems, treating nuclei classically and electrons quantum mechanically.  The basic premise is to take a partial Wigner transform over a set of nuclear degrees of freedom, and then expand the total equation of motion in units of $\hbar.$.  The QCLE includes only terms on the order  of $\hbar^{-1}$ and $\hbar^0=1$; all terms
on the order of $\hbar, \hbar^2, \ldots$ etc. are ignored. 
Formally, the resulting dynamics have some failures -- there is no Jacobi identity and 
correlation functions will not be invariant to time translation\cite{kapral:2001:jcp}. Nevertheless, the dynamics are generally considered to be very accurate. In the context of the spin-boson model, the QCLE is exact.
In this spirit, the QCLE is the underlying
phase space equation against which one would like to compare all other semiclassical approaches\cite{subotnik:2013:qcle_fssh_derive,kapral:2016:chemphys_fssh,kelly:2010:jcp,kelly:2012:map_qcle}.

With this background in mind, recent work has identified a subtle question with regards to nonadiabatic dynamics, namely: Does the QCLE correctly incorporate Berry's phase effects? On the one hand, one might assume that Berry's phase
and Berry's curvature -- both proportional to $\hbar$ --  can arise only if goes beyond the QCLE to include all
 $\hbar^1$ terms in the expansion.  Beside this $\hbar$ expansion argument, note also that Berry's forces are usually derived by considering the Berry potential (or Berry connection) $A \equiv i\hbar \left< \Phi \middle | \nabla \Phi \right>$ of a nearly adiabatic state $\ket{\Phi}$ and, through a gauge transformation acting on the nuclear space, converting the Berry potential to a magnetic force\cite{shankarbook} (just as one changes from the vector potential $A$ to the magnetic field $B$ in electrodynamics\cite{cohentannoudji:photonsandatoms}).   Because gauge
transformations of the classical degrees of freedom are not preserved in a quantum-classical treatment, one might assume that Berry's forces cannot be derived by the QCLE.

On the other hand, recent work by Dou {\em et al.} derived the electronic friction tensor starting from the QCLE\cite{dou:2017:prl} and found the same friction tensor as calculated by a Berry's phase calculation with a complex density matrix
\cite{hedegard:2010:nano_berry,dundas:2012:prb_berry}--suggesting
that Berry's phase should be derivable from the QCLE.  Furthermore,  Berry's phase effects have already been 
isolated and studied within the QCLE for 
real, spin-boson Hamiltonians\cite{izmaylov:2014:qcle_berry} (where the QCLE is exact).
 Thus, in this communication, we seek to tease out the answer to the question: are {\em all} of Berry's phase effects captured by the QCLE, especially for the case of a complex Hamiltonian? Below, we will show clearly that, yes, Berry's phase is derivable from the QCLE through a simple change of representation, as appropriate
in the limit of nearly adiabatic dynamics. We will also show that, while such Berry's phase effects are not captured by surface hopping dynamics, they are captured (at least partially) by Ehrenfest dynamics.

Our conclusions are important for three reasons.  First, because the QCLE has traditionally been regarded as 
the benchmark for all semiclassical algorithms, the  present findings are very reassuring: we may continue
to use the QCLE as the gold standard -- with real or complex Hamiltonians. There is no need to improve
upon the QCLE in the presence of complex Hamiltonians, and in particular we may rest assured that 
the electronic friction tensor as developed in Ref. \onlinecite{dou:2017:prl}
already includes all appropriate Berry's phase effects.
Second, our results should be extremely helpful for understanding and improving upon mixed
quantum classical trajectory techniques. \cite{meek:2016:ci_wf_continuity} Recent work has clearly shown that Tully's fewest switches surface hopping (FSSH) algorithm
 does not include Berry's forces\cite{miao:2019:fssh:complex} for the case of imaginary 
Hamiltonians (though some Berry's phase effects can be captured with FSSH for real Hamiltonians with real conical intersections \cite{izmaylov:2015:fssh_ci_whysowell}).  Even though FSSH is already a partial solution to the QCLE\cite{subotnik:2013:qcle_fssh_derive,kapral:2016:chemphys_fssh},
the failure of surface hopping to recover complex Berry phase effects sheds light on the approximations made in Refs. \onlinecite{subotnik:2013:qcle_fssh_derive} and \onlinecite{kapral:2016:chemphys_fssh},
and thus justifies modifying FSSH to better reproduce the QCLE and treat the case of complex Hamiltonians\cite{miao:2019:fssh:complex}.  At the same time, we can also infer that all approximations to the QCLE based around
Ehrenfest trajectories\cite{meyer_miller:1979,coker:2008:iterative,coker:2012:iterative,kapral:2008:jcp_pbme,kapral:2012:forward_backward,miller:2014:sqc} already include Berry phase effects  and need no such modification.
Third and finally, the present results highlight just how Berry phase arises for nuclear motion
in the adiabatic limit, starting from a very general nonadiabatic approach  but without needing to discuss closed loops in any parameter or function space\cite{aharanov_anandan:prl:1987,pines:1990:arpc:berry}.
Our findings confirm that, at least semiclassically, Berry's phase effects  
can be understood in terms of
well-understood  equations of motion already present in the chemical physics literature
and within all regimes -- from 
the highly nonadiabatic to the highly adiabatic.
With that in mind, we should also be able to learn exactly when Berry's phase is appropriate--what terms must be small in order
to take the semiclassical adiabatic limit?

For convenience below, we will use Einstein summation notation. Electronic states are indexed by 1,2 and nuclear degrees of freedom are indexed by Greek letters ($\alpha, \beta, \gamma$).

\section{Theory}

Without loss of generality, consider the case of two electronic states 1 and 2. 
According to the quantum classical Liouville equation (QCLE)\cite{kapral:1999:jcp}, 
to first order in the electron-nucleus mass ratio $(m/M)^{1/2}$,
the equations of motion for the partial Wigner transform density operator in an adiabatic basis $A^W_{ij}$
are (for the  diagonal and off-diagonal components): 

\begin{eqnarray}\label{kapraldot2}
\frac{\partial}{\partial t}A^W_{11}(R,P,t) 
& =  &  \frac{2 P^{\alpha}}{M^{\alpha}} Re \left(  A^W_{12} d^{\alpha}_{21} \right) 
- \frac{P^{\alpha}}{M^{\alpha}} \frac{\partial A^W_{11}}{\partial R^{\alpha}} 
- F^{\alpha}_{11}  \frac{\partial A^W_{11}}{\partial P^{\alpha}}  
- Re \left( \frac{\partial A^W_{12}}{\partial P^{\alpha}}  F^{\alpha}_{21} \right) 
\end{eqnarray}
and
\begin{eqnarray}\label{kapraldot3}
\frac{\partial}{\partial t}A^W_{12}(R,P,t) &=& \frac{-i}{\hbar} 
\left( V_{11} - V_{22} \right)  A^W_{12}
 - \frac{P^{\alpha}}{M^{\alpha}}  d^{\alpha}_{12} \left( A^W_{22} - A^W_{11}  \right) 
- \frac{P^{\alpha}}{M^{\alpha}} \frac{\partial A^W_{12}}{\partial R^{\alpha}} \\ 
\nonumber
& & 
- \frac{1}{2} \left( F^{\alpha}_{11} + F^{\alpha}_{22} \right) \frac{\partial A^W_{12}}{\partial P^{\alpha}}  
- \frac{1}{2}  F^{\alpha}_{12}
\left( \frac{\partial A^W_{11}}{\partial P^{\alpha}}  +
 \frac{\partial A^W_{22}}{\partial P^{\alpha}}   \right) 
\nonumber
\end{eqnarray}

\noindent A similar equation holds for $A^W_{22}$.  Here, $V_{ii}(\vR)$ are the adiabatic potential energy surfaces and
$\left\{ F_{ij}(\vR) \right\}$ are the  set of forces,
$F^{\alpha}_{ij}(\vR) \equiv  -\left<\Phi_i(\vR) | \frac{\partial V}{\partial R^{\alpha}} | \Phi_j(\vR) \right> $.
\{$\ket{\Phi_i(\vR)}$\} are an adiabatic basis set of electronic states, and $d^{\alpha}_{ij}(\vR)$ are the derivative couplings, 
$d^{\alpha}_{ij}(\vR) \equiv  F^{\alpha}_{ij}(\vR) /\left(V_{ii}(\vR) - V_{jj}(\vR) \right)$. We note that $d_{ij} = -d_{ji}^*$

At this point, we assume that all dynamics are being propagated near the adiabatic limit, with the population
of state 1 close to unity (and only barely changing in time).
Thus, the coherences are nearly stationary and (hopefully) not evolving much as well.  In such a case,
we can identify the steady state equation of motion for the coherences in Eq. \ref{kapraldot3} by 
ignoring any evolution of the coherences:\cite{dou:2017:prl}

\begin{eqnarray}
\frac{-i}{\hbar}\left( V_{11} - V_{22} \right)  A^W_{12}
 - \frac{P^{\alpha}}{M^{\alpha}}  d^{\alpha}_{12} \left( A^W_{22} - A^W_{11}  \right)  
- \frac{1}{2}  F^{\alpha}_{12}
\left( \frac{\partial A^W_{11}}{\partial P^{\alpha}}  +
 \frac{\partial A^W_{22}}{\partial P^{\alpha}}   \right) 
= 0,
\end{eqnarray}

\noindent which has the solution $A^{W}_{12} = \zeta$, where 
\begin{eqnarray}
\zeta &\equiv& 
\frac{i \hbar \frac{P^{\gamma}}{M^{\gamma}}  d^{\gamma}_{12} \left( A^W_{22} - A^W_{11}  \right) }{V_{11} - V_{22}} + \frac{i \hbar}{2} 
d^{\alpha}_{12}
\left( \frac{\partial A^W_{11}}{\partial P^{\alpha}}  +
 \frac{\partial A^W_{22}}{\partial P^{\alpha}}   \right) 
\end{eqnarray}

Thereafter, we change variables from $A^{W}_{12}$ to 
$B^W_{12} \equiv A^W_{12} - \zeta$.  The equations of motion for the populations are modified as follows:

\begin{eqnarray}\label{eq_A11}
\frac{\partial}{\partial t}A^W_{11}(R,P,t) 
& =  &  \frac{2 P^{\alpha}}{M^{\alpha}} Re \left(  \left( B^W_{12} + \zeta \right) d^{\alpha}_{21} \right) - \frac{P^{\alpha}}{M^{\alpha}} \frac{\partial A^W_{11}}{\partial R^{\alpha}} - F^{\alpha}_{11}  \frac{\partial A^W_{11}}{\partial P^{\alpha}}  - Re \left( \frac{\partial \left( B^W_{12} + \zeta \right)}{\partial P^{\alpha}}  F^{\alpha}_{21} \right)  \nonumber \\ 
& =  &  \frac{2 P^{\alpha}}{M^{\alpha}} Re \left(   B^W_{12}  d^{\alpha}_{21} \right) - \frac{P^{\alpha}}{M^{\alpha}} \frac{\partial A^W_{11}}{\partial R^{\alpha}} - F^{\alpha}_{11}  \frac{\partial A^W_{11}}{\partial P^{\alpha}}  - Re \left( \frac{\partial  B^W_{12} }{\partial P^{\alpha}}  F^{\alpha}_{21} \right)  + 2 \hbar Im \left(  d^{\beta}_{21} \frac{P^{\alpha}}{M^{\alpha}}  d^{\alpha}_{12}    \right) \frac{\partial  A^W_{11} }{\partial P^{\beta}} 
\nonumber  \\
\end{eqnarray}

The equations of motion for the coherences are more involved and given in Appendix \ref{apdx_B12}.
If we assume that we are in the adiabatic limit moving along adiabat 1, noting $B^W_{12}$ vanishes in the adiabatic limit,
Eq. \ref{eq_A11} simplifies:
\begin{eqnarray}
\frac{\partial}{\partial t}A^W_{11}(R,P,t) 
 =   - \frac{P^{\alpha}}{M^{\alpha}} \frac{\partial A^W_{11}}{\partial R^{\alpha}} - F^{\alpha}_{11}  \frac{\partial A^W_{11}}{\partial P^{\alpha}}   + 2 \hbar Im \left(  d^{\beta}_{21} \frac{P^{\alpha}}{M^{\alpha}}  d^{\alpha}_{12}    \right) \frac{\partial  A^W_{11} }{\partial P^{\beta}}   \nonumber 
\end{eqnarray}
The total effective force is the usual adiabatic force $\vec{F}_{11}$ plus
the Berry magnetic  force 
\begin{eqnarray}\label{eq_berry_force}
\vec{F}^{B}_{11} =  - 2 \hbar Im \left(   \vec{d}_{21} \frac{P^{\alpha}}{M^{\alpha}}  d^{\alpha}_{12}      \right)
=  2 \hbar Im \left(   \vec{d}_{12} \frac{P^{\alpha}}{M^{\alpha}}  d^{\alpha}_{21}      \right)
\end{eqnarray}
which arises as the curl of the Berry connection and which vanishes for a real Hamiltonian.  Clearly, the QCLE already includes the effects of Berry phase.

\section{Discussion: Implications for Semiclassical Dynamics}

Having successfully isolated Berry's phase within the QCLE, let us now discuss
the implications of our findings for mixed quantum classical methods. After all,
one can view semiclassical nonadiabatic dynamics methods as approximations
to the QCLE, and so one must wonder: do the standard semiclassical approaches  (surface hopping
and Ehrenfest dynamics) also account for Berry's phase?

Consider the Hamiltonian that was introduced in Ref. \onlinecite{miao:2019:fssh:complex}:

\begin{equation} \label{eq_Hamiltonian}
\begin{aligned} 
H & = A\begin{bmatrix}
-\cos{\theta}& \sin{\theta} e^{i\phi} \\ 
\sin{\theta} e^{-i\phi} & \cos{\theta} 
\end{bmatrix} \\
\end{aligned}
\end{equation}
\noindent where 
$\theta(x) \equiv \frac{\pi}{2} \left(erf(Bx) + 1\right)$, and 
$\phi(y) \equiv Wy$. 

For this Hamiltonian, the adiabatic surfaces are completely flat. For an incoming wavepacket on surface 2
 beginning at $x = -\infty$ and traveling in the $+x$ direction, the exact solution predicts that the wavepacket should bend upwards.  If W is small enough,
the asymptotic momentum of the transmitted wavepacket should be W.  This behavior follows by considering Berry's force. 
For the  Hamiltonian in Eq. \ref{eq_Hamiltonian}, the Berry force is
$\vec{F}^B_2 = 2\hbar Im\left[\vec{d}_{21}(\frac{\vec{P}}{M}\cdot\vec{d}_{12})\right] = \frac{\hbar W}{2} \partial_x \theta \sin{\theta} (-\frac{P^y}{M}, \frac{P^x}{M})$. 
When $W$ is small enough, we can assume that $P^x$ is roughly constant, and so we may calculate the final $y$-direction momentum (at the end of  a scattering event) by integrating the $y$-component of the Berry force:
\begin{equation}
\begin{aligned}
p^y &= \int_0^{\infty} \frac{\hbar W}{2} \partial_x \theta \sin{\theta} \frac{P^x}{M} dt = \hbar W
\end{aligned}
\end{equation}
Of course, if W is not small, the result above is invalid; instead, the wavepacket can actually split apart
and  a portion of the wavepacket will reflect -- even though the adiabats are completely flat. 

Now, the example above makes very clear (as shown in Ref. \onlinecite{miao:2019:fssh:complex})
that the FSSH algorithm does not capture Berry's phase effects in the case of a {\em complex} Hamiltonian\cite{izmaylov:2014:qcle_berry}.  FSSH dictates 
motion along adiabats and the algorithm
will not predict any bending or reflection; for this reason, in Ref. \onlinecite{miao:2019:fssh:complex}, we have recommended augmenting FSSH dynamics with the Berry force $\vec{F}_B$ 
(in Eq. \ref{eq_berry_force}) in order to better agree with the QCLE and capture  the correct quantum dynamics. Clearly, further benchmarking of such a corrected FSSH approach will be necessary.

At this point,  it is worthwhile to consider the natural alternative to FSSH dynamics, namely Ehrenfest dynamics. 
Does Ehrenfest dynamics correctly account for Berry's phase, or does it also require a Berry phase correction?
We will now argue (analytically and numerically) that  Ehrenfest 
dynamics {\em do already include}  Berry's phase;  for the Hamiltonian in Eq. \ref{eq_Hamiltonian},
 in the limit of nearly adiabatic
dynamics, Ehrenfest trajectories will bend the correct amount. Thus, 
despite the many failures of Ehrenfest 
dynamics (i.e. a lack of branching\cite{tully:fssh}, a lack of detailed balance\cite{tully:2005:detailedbalance,tully:2008:detailedbalance}, a lack of decoherence \cite{subotnik:2010:augehr,schwartz:mfsd1, schwartz:mfsd2,truhlar:2004:surfacehop2,truhlar:review:surfacehop}),
a correction  for Berry's phase effects is not needed.

To prove this point, consider the propagation of the wave function during an Ehrenfest trajectory
for the Hamiltonian in Eq. \ref{eq_Hamiltonian}:
\begin{equation}
\begin{aligned}
\dot{c}_1 &= -\frac{iE_1}{\hbar} c_1 - \frac{\vec{P}}{M}\cdot \vec{d}_{12} c_2 \\
\dot{c}_2 &= -\frac{iE_2}{\hbar} c_2 - \frac{\vec{P}}{M}\cdot \vec{d}_{21} c_1 \\
\end{aligned}
\end{equation}
The time evolution of density matrix element ($\rho_{jk}\equiv c_jc_k^*$) is  
\begin{equation}
\begin{aligned}
\dot{\rho_{21}} = i\omega_{12}\rho_{21} + \left(\frac{\vec{P}}{M}\cdot \vec{d}_{21} \right)(\rho_{22} - \rho_{11}) \approx  i\omega_{12}\rho_{21} + \left(\frac{\vec{P}}{M}\cdot \vec{d}_{21} \right)
\end{aligned}
\end{equation}
Here $\omega_{12} \equiv (E_1 - E_2) / \hbar$ and the adiabatic limit has been invoked such that trajectories are moving along surface 2  at all times ($\rho_{22} \approx 1$).
Solving the above ODE with initial condition $\rho_{21}(0) = 0$, we find:
\begin{equation}
\begin{aligned}
\rho_{21}(t) = e^{i\omega_{12}t} \int_0^t \left(\frac{\vec{P}}{M}\cdot \vec{d}_{21} \right) e^{-i\omega_{12}\tau} d\tau
\end{aligned}
\end{equation}
For the Hamiltonian in Eq. \ref{eq_Hamiltonian}, one can compute $\vec{d}_{21} = \frac{1}{2}\left(-\partial_x\theta, iW\sin{\theta}\right)$. Thus, for the small $W$ case where $P^x$ is constant and $P^y \approx 0$, we can
integrate $\rho_{21}(t)$ by parts (with the fact that $\left.\frac{\partial^k \theta}{\partial t^k}\right|_{t=0} = 0$ for any order of $k$ assuming that we initially start far away from the crossing):
\begin{equation}
\begin{aligned}
\rho_{21}(t)
&=  -\frac{1}{2} e^{i\omega_{12}t} \int_0^t \frac{\partial \theta}{\partial \tau} e^{-i\omega_{12}\tau} d\tau \\
    &= -\frac{1}{2}e^{i\omega_{12}t} \left(-\frac{1}{i\omega_{12}} \left.e^{-i\omega_{12}\tau} \frac{\partial \theta(\tau)}{\partial \tau}\right|_0^t + \frac{1}{i\omega_{12}} \int_0^t \frac{\partial^2 \theta(\tau)}{\partial \tau^2} e^{-i\omega_{12}\tau}d\tau \right) \\
&=\frac{1}{2}\sum_{k=1}^{\infty}\frac{1}{(i\omega_{12})^k} \frac{\partial^k \theta}{\partial t^k}
\end{aligned}
\end{equation}
To approximate the above series, we use the definition of $\theta$ in Eq. \ref{eq_Hamiltonian}. 
For the term associated with $\frac{\partial^{k}\theta}{\partial t^k}$, one can show that the order of magnitude is $(\frac{B P^x}{\omega_{12} M})^k$. Thus, if $\frac{B P^x}{\omega_{12} M}$ is small, the first term will dominate the series, and the average force (as well as the final momentum) can be calculated as 
\begin{equation}
\begin{aligned}
\avg{F^y}(t) 
& = 2Re\left(F_{12}(t)\rho_{21}(t)\right) 
= \frac{\hbar W}{2} \sin{\theta(t)} \frac{\partial \theta}{\partial t} \\
\end{aligned}
\end{equation}
\begin{equation}\label{eq_ehrenfest_py}
\begin{aligned}
p^y 
&= \int_0^\infty \avg{F^y}(t)dt = \hbar W
\end{aligned}
\end{equation}
From this argument, it is clear that Berry phase effects are already included in Ehrenfest dynamics (unlike FSSH) and there is no need for any additional corrections.

\begin{figure}[h!]
	\centering
	\includegraphics[width=6in]{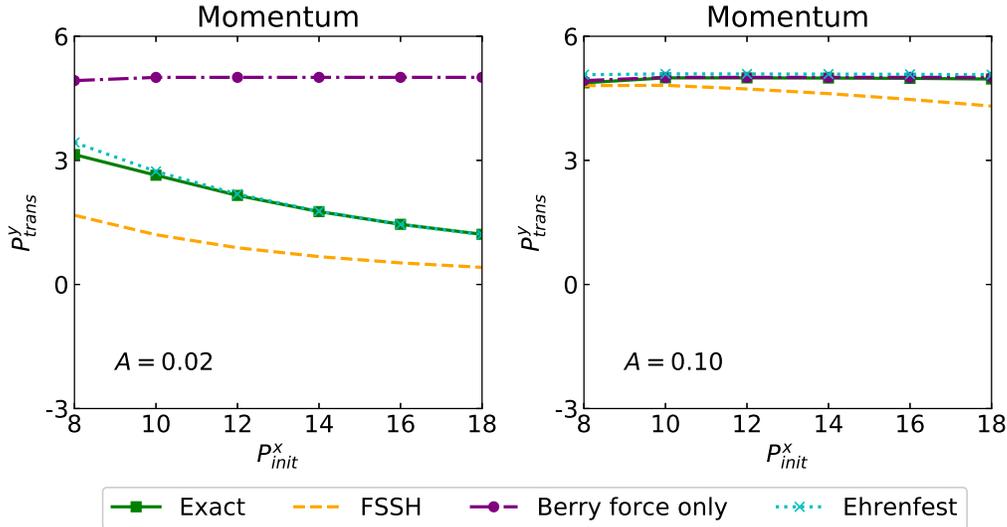}
	\caption{
		Transmitted and average $y$-momentum as a function of initial momentum $P^x_{init}$.
		Left: $A = 0.02$, corresponding to diabatic regime;
		right: $A = 0.10$, corresponding to adiabatic regime. 
		$P^y_{init} = 0$.
		Little reflection is observed for both cases.
		While FSSH (with an imposed Berry force) captures the correct trend, it is numerically outperformed by Ehrenfest dynamics (which naturally accounts for Berry's phase). Adiabatic dynamics with a Berry force works only in the adiabatic regime, $A = 0.1$.
	}
	\label{fig_norefl}
\end{figure}

%
Finally, in order to numerically assess the relative value of Berry-corrected FSSH and Ehrenfest dynamics, in Fig. \ref{fig_norefl} we plot the transmitted $y$ momentum as a function of incoming momentum in both the adiabatic and diabatic regimes for the Hamiltonian in Eq. \ref{eq_Hamiltonian}. For this data set, we set $W = 5$, $M = 1000$, and $B = 3.0$,
and as far as FSSH is concerned, we rescale all velocities in the $x-$direction whenever a hop occurs. We imagine a particle coming on adiabat $2$  from the left. For comparison, besides Ehrenfest and FSSH, we also plot results for exact dynamics as well as classical adiabatic dynamics with Berry's forces. Reflection is rare and not important here. As one can see from the figure, Ehrenfest outperforms Berry-corrected FSSH in both the diabatic and adiabatic regimes as far as average momentum, indicating that Ehrenfest dynamics work better than FSSH even after a Berry-phase correction.  Clearly, despite its many failures\cite{ tully:fssh,
tully:2005:detailedbalance,tully:2008:detailedbalance,
subotnik:2010:augehr,schwartz:mfsd1, schwartz:mfsd2,truhlar:2004:surfacehop2,truhlar:review:surfacehop}, 
Ehrenfest dynamics incorporate Berry's forces naturally and work very well for this problem
of flat adiabatic surfaces; FSSH captures the correct trends but has a relatively larger error. 

Lastly, using a Berry force and running purely adiabatic dynamics can be very accurate in the adiabatic regime, i.e. $A = 0.10$.  That being said, running adiabatic dynamics with a Berry force is awful in the diabatic regime, i.e. $A = 0.02$.



\section{Conclusions}
In this communication, we have demonstrated that, even though the QCLE arises from a cut-off in $\hbar$ at order 0 from the Wigner distribution equation of motion, QCLE dynamics {\em do} include Berry's phase effects (which are of order $\hbar$).  As such, 
even though Berry's phase effects are not usually\cite{izmaylov:2014:qcle_berry} studied explicitly with the QCLE, 
if classical nuclei are sufficient, one can safely study many physical problems with geometric phase using the well-established QCLE
and approximations thereof; of course, the bigger problem remains how to solve the QCLE
in practice.  Here, we have shown that Tully's surface hopping approximation to the QCLE
does not include Berry's phase effects (when the Hamiltonian is complex),
and we have recently made the sensible suggestion to simply add in the Berry force (Eq. \ref{eq_berry_force})\cite{miao:2019:fssh:complex}. At the same time,  we have also shown that Ehrenfest dynamics do contain Berry's phase and, as such, no extra force is required. 

Looking forward, the keen reader should observe that our 
model problem  here (Eq. \ref{eq_Hamiltonian}) is an extremely
unphysical example whereby
one can easily isolate Berry's phase effects. For most problems with avoided crossings and
 conical intersections\cite{levine:2007:ci_isomerization}, the adiabatic force difference will not be constant and surface hopping is usually expected to be a better approximation than Ehrenfest dynamics
at recovering long time dynamics
(e.g. populations during electron transfer dynamics\cite{landry:2013:electronicproperties}).  Further research will need to assess whether FSSH can still be improved and how to incorporate decoherence\cite{rossky:1991:surfacehop,rossky:1994:surfacehop, schwartz:1996:solve, rossky:2002:surfacehop1,rossky:2002:surfacehop2,prezhdo:1997:surfacehop,
prezhdo:1999:surfacehop,shs:1999:jcp,truhlar:2000:surfacehop, truhlar:2001:surfacehop, truhlar:2001:surfacehop2,
truhlar:2004:surfacehop, truhlar:2005:surfacehop} within a Berry-force modified algorithm. Another important question is how
to choose a momentum rescaling direction for surface hopping; here, for the Hamiltonian in Eq. \ref{eq_Hamiltonian}, we simply chose $x$ as the rescaling direction but for a more general Hamiltonian, a better ansatz is needed.  Unfortunately, preliminary evidence suggests that the algorithms in Ref. \onlinecite{miao:2019:fssh:complex} are not yet optimal; perhaps the different form of the QCLE (as present in Eqs. \ref{eq_A11} and \ref{eq_B12})
will be useful for future derivations. At the very least, the equations should yield insight
into exactly when one can make the adiabatic approximation and ignore $B_{12}$.

Finally and most importantly, now that we know that Berry's phase dynamical effects are already
 included within the QCLE,
this communication raises the distinct possibility of using the QCLE (and approximations thereof) to study
coupled nuclear-electronic motion on the surfaces of topological materials, where the electronic Hamiltonian is complex and electronic Berry's phase effects are already known
to be of crucial importance\cite{vanderbilt:2009:prl}.   One must wonder if one will 
learn something new about nonadiabatic dynamics in such a context.

\section{Supplementary Material}
See supplementary material for the detailed derivations of $\frac{\partial A^W_{11}}{\partial t}$ (Eq. \ref{eq_A11}) and $\frac{\partial B^W_{12}}{\partial t}$ (Eq. \ref{eq_B12}). 

\section{Acknowledgement} 
    This material is based upon work supported by the National Science Foundation under Grant No. CHE-1764365. 
    J.E.S. acknowledges a David and Lucille Packard Fellowship.
    J.E.S. thanks Abe Nitzan, Hsing-Ta Chen, Tao E. Li, Zeyu Zhou, Alec Coffman and Zuxin Jin for interesting conversations.

\appendix
\section{Expression for $\frac{\partial B^W_{12}}{\partial t}$} \label{apdx_B12}
After a great deal of algebra (see the Supporting Information), the equation of motion for $B^W_{12}$ can be shown to be:
\begin{eqnarray}\label{eq_B12}
\frac{\partial}{\partial t}B^W_{12}(R,P,t) 
& = & \frac{-i}{\hbar} 
\left( V_{11} - V_{22} \right)   B^W_{12} 
- \frac{P^{\alpha}}{M^{\alpha}} \frac{\partial  B^W_{12} }{\partial R^{\alpha}} 
- \frac{1}{2} \left( F^{\alpha}_{11} + F^{\alpha}_{22} \right) \frac{\partial B^W_{12} }{\partial P^{\alpha}}  
\\
\nonumber
& & - i \hbar \frac{P^{\alpha}}{M^{\alpha}} \frac{\partial  d_{12}^{\gamma} }{\partial R^{\alpha}}
\frac{P^{\gamma}}{M^{\gamma}} \frac{  \left( A^W_{22} - A^W_{11}  \right) }{\left(V_{11} - V_{22}\right)} \\
\nonumber
& & - i \hbar \frac{P^{\alpha}}{M^{\alpha}} 
\frac{d_{12}^{\gamma}P^{\gamma}}{M^{\gamma}} \frac{  \left( A^W_{22} - A^W_{11}  \right) }{\left( V_{11} - V_{22} \right)^2} \left(F_{11}^{\alpha} - F_{22}^{\alpha}\right) \\
\nonumber
& & - \frac{i \hbar  P^{\alpha}}{2 M^{\alpha}} 
\frac{\partial d_{12}^{\gamma}}{\partial R^{\alpha}}
\left(\frac{\partial   A^W_{11} }{\partial P^{\gamma}} + \frac{\partial   A^W_{22}}{\partial P^{\gamma}}  
 \right)  \\
\nonumber
& & - \frac{i \hbar}{2} \left( F^{\alpha}_{11} + F^{\alpha}_{22} \right) 
\frac{  d^{\alpha}_{12}  \left( A^W_{22} - A^W_{11} \right)      }{
M^{\alpha} \left( V_{11} - V_{22} \right) }   \\ 
\nonumber
& &
+\frac{i \hbar}{4} d_{12}^{\gamma} 
\left( (F_{22}^{\alpha} - F_{11}^{\alpha}) \left(\frac{\partial^2 A^W_{22}}{\partial P^{\alpha} \partial P^{\gamma}} - \frac{\partial^2 A^W_{11}}{\partial P^{\alpha} \partial P^{\gamma}}\right) \right) \\ 
\nonumber
& & 
+ \frac{i \hbar P^{\gamma}  d^{\gamma}_{12}}
{M^{\gamma} \left( V_{11} - V_{22} \right) }
\left( \frac{4 P^{\alpha}}{M^{\alpha}} Re \left(   B^W_{12}  d^{\alpha}_{21} \right) + 
\frac{1}{2}\left( F^{\alpha}_{22}  - F^{\alpha}_{11}\right) \left( \frac{\partial A^W_{22}}{\partial P^{\alpha}}  -  \frac{\partial A^W_{11}}{\partial P^{\alpha}} \right) \right)\\
\nonumber
& & + \frac{i \hbar}{2} \frac{d_{12}^{\alpha}}{M^{\alpha}} 
\left( \frac{\partial A^W_{11}}{\partial R^{\alpha}}
      +\frac{\partial A^W_{22}}{\partial R^{\alpha}} \right) \\
& & + i\hbar  d_{12}^{\gamma}  Re\left( \frac{\partial^2 B^W_{12}}{\partial  P^{\alpha} \partial P^{\gamma}} F_{21}^{\alpha} \right) \\
\nonumber & &
+ \frac{i\hbar P^{\gamma}d^{\gamma}_{12}}{M^{\gamma}(V_{11}-V_{22})} \left[2\hbar Im\left(d^{\alpha}_{21}\frac{P^{\beta}}{M^{\beta}}d^{\beta}_{12}\right)\left(\frac{\partial A^W_{11}}{\partial P^{\alpha}} + \frac{\partial A^W_{22}}{\partial P^{\alpha}}\right)
\right] 
\\ \nonumber  & &
- i\hbar d^{\gamma}_{12} \left[\hbar Im\left( \frac{ d^{\alpha}_{12} d^{\gamma}_{21}}{M^{\gamma}} \right) \left( \frac{\partial A^W_{22}}{\partial P^{\alpha}}  -  \frac{\partial A^W_{11}}{\partial P^{\alpha}} \right) \right] 
\\ \nonumber  & &
- i\hbar d^{\gamma}_{12} \left[\hbar Im\left( d^{\alpha}_{12} \frac{P^{\beta}}{M^{\beta}}  d^{\beta}_{21}\right) \left(\frac{\partial^2 A^W_{22}}{\partial P^{\alpha} \partial P^{\gamma}} - \frac{\partial^2 A^W_{11}}{\partial P^{\alpha} \partial P^{\gamma}}\right) \right] 
\end{eqnarray}


\bibliographystyle{apsrev}

\end{document}


\title{Supporting Information}
	\maketitle

	\section{Deriving Berry Force from QCLE}
	
	\subsection{The $\frac{\partial A^W_{11}}{\partial t}$ Expression }
	To derive the $\frac{\partial A^W_{11}}{\partial t}$ expression (Eq. 5), we start from the definition of $\zeta$ (Eq. 4) in the paper
	\begin{eqnarray}
	\zeta 
	& = &  \frac{i \hbar \Delta}{V_{11}-V_{22}}
	\frac{P^{\gamma}}{M^{\gamma}}  d^{\gamma}_{12} +  \frac{i \hbar}{2} 
	d^{\alpha}_{12} \frac{\partial \Gamma}{\partial P^{\alpha}} \\
	\Delta & \equiv & A^W_{22} - A^W_{11} \\ 
	\Gamma & \equiv & A^W_{22} + A^W_{11} \\  \label{eq_Bdef}
	B^W_{12} &\equiv& A^W_{12} - \zeta 
	\end{eqnarray}
	Using \refeq{eq_Bdef}, Eq. 1 in the paper becomes
	\begin{eqnarray}
	\frac{\partial A^W_{11}}{\partial t}
	& =  &  \frac{2 P^{\alpha}}{M^{\alpha}} Re \left(  \left( B^W_{12} + \zeta \right) d^{\alpha}_{21} \right) - \frac{P^{\alpha}}{M^{\alpha}} \frac{\partial A^W_{11}}{\partial R^{\alpha}} - F^{\alpha}_{11}  \frac{\partial A^W_{11}}{\partial P^{\alpha}}  - Re \left( \frac{\partial \left( B^W_{12} + \zeta \right)}{\partial P^{\alpha}}  F^{\alpha}_{21} \right)  
	\\ \nonumber
	& =  &  \frac{2 P^{\alpha}}{M^{\alpha}} Re \left(B^W_{12} d^{\alpha}_{21} \right) - \frac{P^{\alpha}}{M^{\alpha}} \frac{\partial A^W_{11}}{\partial R^{\alpha}} - F^{\alpha}_{11}  \frac{\partial A^W_{11}}{\partial P^{\alpha}}  - Re \left( \frac{\partial  B^W_{12} }{\partial P^{\alpha}}  F^{\alpha}_{21} \right) 
	 \\ \nonumber
	& & + \frac{2 P^{\alpha}}{M^{\alpha}} Re \left(\zeta d^{\alpha}_{21} \right) - Re \left( \frac{\partial  \zeta }{\partial P^{\alpha}}  F^{\alpha}_{21} \right) 
	\end{eqnarray}
	We focus on the last 2 terms above:
	\begin{eqnarray}\label{eq_A11_term1}
	\frac{2 P^{\alpha}}{M^{\alpha}} Re \left(\zeta d^{\alpha}_{21} \right) 
	&=&  Re\left(\frac{i \hbar \Delta}{V_{11}-V_{22}}
	\frac{P^{\gamma}}{M^{\gamma}}  d^{\gamma}_{12}\frac{P^{\alpha}}{M^{\alpha}}d^{\alpha}_{21}\right)
	 + Re \left(\frac{i \hbar}{2}d^{\gamma}_{12} \frac{\partial \Gamma}{\partial P^{\gamma}} \frac{P^{\alpha}}{M^{\alpha}}d^{\alpha}_{21}\right) 
	 \\ \nonumber
	 & = & 0 + Re \left(\frac{i\hbar}{2}d^{\gamma}_{12}\frac{P^{\alpha}}{M^{\alpha}}d^{\alpha}_{21} \right)\frac{\partial \Gamma}{\partial P^{\gamma}} \\\nonumber
	 & = & \hbar Im\left(d^{\gamma}_{21}\frac{P^{\alpha}}{M^{\alpha}}d^{\alpha}_{12}\right)\frac{\partial \Gamma}{\partial P^{\gamma}} \\\nonumber
	 & = & \hbar Im\left(d^{\alpha}_{21}\frac{P^{\gamma}}{M^{\gamma}}d^{\gamma}_{12}\right)\frac{\partial \Gamma}{\partial P^{\alpha}} \\\nonumber
	\end{eqnarray}
	For the last term:
	\begin{eqnarray}\label{eq_A11_term2}
	 - Re \left( \frac{\partial  \zeta }{\partial P^{\alpha}}  F^{\alpha}_{21} \right) 
	 &=& 
	 -\frac{1}{2} Re\left(-i \hbar \frac{P^{\gamma}}{M^{\gamma}}  d^{\gamma}_{12} \frac{\partial \Delta}{\partial P^{\alpha}} d^{\alpha}_{21} \right) \\\nonumber
	 & & -\frac{1}{2} Re\left(-i \hbar\Delta \frac{d^{\alpha}_{12}  d^{\alpha}_{21}}{M^{\alpha}} \right)  \\\nonumber
	 & & -\frac{1}{2} Re\left(\frac{i \hbar}{2} d^{\gamma}_{12} \frac{\partial^2 \Gamma}{\partial P^{\gamma}\partial P^{\alpha}}F^{\alpha}_{21} \right)\\\nonumber
	 &=&
	 -\frac{1}{2} Re\left( i\hbar d^{\alpha}_{12} \frac{P^{\gamma}}{M^{\gamma}}  d^{\gamma}_{21}\right)\frac{\partial \Delta}{\partial P^{\alpha}} 
	  + 0 + 0\\\nonumber
	 &=& \hbar Im\left( d^{\alpha}_{12} \frac{P^{\gamma}}{M^{\gamma}}  d^{\gamma}_{21}\right) \frac{\partial \Delta}{\partial P^{\alpha}} 
	\end{eqnarray}
	Combining the two terms, we find:
	\begin{eqnarray}
	\frac{2 P^{\alpha}}{M^{\alpha}} Re \left(\zeta d^{\alpha}_{21} \right) - Re \left( \frac{\partial  \zeta }{\partial P^{\alpha}}  F^{\alpha}_{21} \right) 
	&=&
	\hbar Im\left(d^{\alpha}_{21}\frac{P^{\gamma}}{M^{\gamma}}d^{\gamma}_{12}\right)\frac{\partial \Gamma}{\partial P^{\alpha}} 
	+  \hbar Im\left( d^{\alpha}_{12} \frac{P^{\gamma}}{M^{\gamma}}  d^{\gamma}_{21}\right) \frac{\partial \Delta}{\partial P^{\alpha}} \\\nonumber
	&=& \hbar Im\left( d^{\alpha}_{12} \frac{P^{\gamma}}{M^{\gamma}}  d^{\gamma}_{21}\right) \frac{\partial (\Delta-\Gamma)}{\partial P^{\alpha}} \\\nonumber
	&=& -2\hbar Im\left( d^{\alpha}_{12} \frac{P^{\gamma}}{M^{\gamma}}  d^{\gamma}_{21}\right) \frac{\partial A^W_{11}}{\partial P^{\alpha}} \\\nonumber
	&=& 2\hbar Im\left( d^{\gamma}_{21} \frac{P^{\alpha}}{M^{\alpha}}  d^{\alpha}_{12}\right) \frac{\partial A^W_{11}}{\partial P^{\gamma}} 
	\end{eqnarray}
	Therefore, we find:
	\begin{eqnarray}
	\frac{\partial A^W_{11}}{\partial t}
	& =  &  \frac{2 P^{\alpha}}{M^{\alpha}} Re \left(B^W_{12} d^{\alpha}_{21} \right) - \frac{P^{\alpha}}{M^{\alpha}} \frac{\partial A^W_{11}}{\partial R^{\alpha}} - F^{\alpha}_{11}  \frac{\partial A^W_{11}}{\partial P^{\alpha}}  - Re \left( \frac{\partial  B^W_{12} }{\partial P^{\alpha}}  F^{\alpha}_{21} \right)  \\\nonumber
	& & + 2\hbar Im\left( d^{\gamma}_{21} \frac{P^{\alpha}}{M^{\alpha}}  d^{\alpha}_{12}\right) \frac{\partial A^W_{11}}{\partial P^{\gamma}}
	\end{eqnarray}
	This is the $A^W_{11}$ result. 
	
	\subsection{The $\frac{\partial B^W_{12}}{\partial t}$ Expression }
	Now let's consider $B^W_{12} \equiv A^W_{12} - \zeta$:
	\begin{eqnarray} \label{eq_B12_expansion}
	\frac{\partial B^W_{12} }{\partial t}
	&=& \frac{\partial A^W_{12} }{\partial t} - \frac{\partial \zeta }{\partial t} \\\nonumber
	&=& \frac{-i}{\hbar} 
	\left( V_{11} - V_{22} \right)  A^W_{12}
	- \frac{P^{\alpha}}{M^{\alpha}}  d^{\alpha}_{12} \Delta
	- \frac{P^{\alpha}}{M^{\alpha}} \frac{\partial A^W_{12}}{\partial R^{\alpha}} \\ \nonumber
	& & 
	- \frac{1}{2} \left( F^{\alpha}_{11} + F^{\alpha}_{22} \right) \frac{\partial A^W_{12}}{\partial P^{\alpha}}  
	- \frac{1}{2}  F^{\alpha}_{12} \frac{\partial \Gamma}{\partial P^{\alpha}} 
	- \frac{\partial \zeta }{\partial t}
	\nonumber \\\nonumber
	&=& -\frac{i}{\hbar} 
	\left( V_{11} - V_{22} \right)  B^W_{12}
	- \frac{P^{\alpha}}{M^{\alpha}}  d^{\alpha}_{12} \Delta
	- \frac{P^{\alpha}}{M^{\alpha}} \frac{\partial B^W_{12}}{\partial R^{\alpha}} \\ \nonumber
	\nonumber
	& & 
	- \frac{1}{2} \left( F^{\alpha}_{11} + F^{\alpha}_{22} \right) \frac{\partial B^W_{12}}{\partial P^{\alpha}}  
	- \frac{1}{2}  F^{\alpha}_{12} \frac{\partial \Gamma}{\partial P^{\alpha}}  \\
	\nonumber
	& &
	-\frac{i}{\hbar} \left( V_{11} - V_{22} \right)  \zeta
	- \frac{P^{\alpha}}{M^{\alpha}} \frac{\partial \zeta}{\partial R^{\alpha}} 
	- \frac{1}{2} \left( F^{\alpha}_{11} + F^{\alpha}_{22} \right) \frac{\partial\zeta }{\partial P^{\alpha}}  
	- \frac{\partial \zeta }{\partial t} 
	\nonumber
	\end{eqnarray}
	From the definition of $\zeta$ we have
	\begin{eqnarray}
	-\frac{i}{\hbar}(V_{11} - V_{22})\zeta - \frac{P^{\alpha}}{M^{\alpha}}d^{\alpha}_{12}\Delta - \frac{1}{2}F_{12}^{\alpha} \frac{\partial \Gamma}{\partial P^{\alpha}} = 0
	\end{eqnarray}
	and so three terms vanish in \refeq{eq_B12_expansion}, leading to: 
	\begin{eqnarray} \label{eq_B12_simple}
	\frac{\partial B^W_{12} }{\partial t}
	&=& -\frac{i}{\hbar} 
	\left( V_{11} - V_{22} \right)  B^W_{12}
	- \frac{P^{\alpha}}{M^{\alpha}} \frac{\partial B^W_{12}}{\partial R^{\alpha}} 
	- \frac{1}{2} \left( F^{\alpha}_{11} + F^{\alpha}_{22} \right) \frac{\partial B^W_{12}}{\partial P^{\alpha}} \\
	\nonumber
	& &
	- \frac{1}{2} \left( F^{\alpha}_{11} + F^{\alpha}_{22} \right) \frac{\partial\zeta }{\partial P^{\alpha}}  
	- \frac{P^{\alpha}}{M^{\alpha}} \frac{\partial \zeta}{\partial R^{\alpha}} 
	- \frac{\partial \zeta }{\partial t} 
	\nonumber
	\end{eqnarray}
	We must now evaluate the last three terms in \refeq{eq_B12_simple}\\
   The first term is
	\begin{eqnarray}
	- \frac{1}{2} \left( F^{\alpha}_{11} + F^{\alpha}_{22} \right) \frac{\partial\zeta }{\partial P^{\alpha}}  
	&=& - \frac{1}{2} \left( F^{\alpha}_{11} + F^{\alpha}_{22} \right) 
	\left(\frac{i \hbar}{V_{11}-V_{22}}\frac{\partial \Delta}{\partial P^{\alpha} }
	\frac{P^{\gamma}}{M^{\gamma}}  d^{\gamma}_{12} 
	+ \frac{i \hbar \Delta}{V_{11}-V_{22}} 
	\frac{d^{\alpha}_{12}}{M^{\alpha}}  
	+  \frac{i \hbar}{2} 
	d^{\gamma}_{12} \frac{\partial^2 \Gamma}{\partial P^{\gamma} \partial P^{\alpha}}
	\right)  \\
	\nonumber
	&=&  
	- \frac{i \hbar}{2} (F^{\alpha}_{11} + F^{\alpha}_{22})
	\frac{P^{\gamma}d^{\gamma}_{12}}{M^{\gamma}(V_{11}-V_{22})} \frac{\partial \Delta}{\partial P^{\alpha} } 
	\\ \nonumber & &
	- \frac{i \hbar}{2}(F^{\alpha}_{11} + F^{\alpha}_{22})
	\frac{d^{\alpha}_{12} \Delta}{M^{\alpha}(V_{11}-V_{22})} 
	\\ \nonumber & &
	- \frac{i \hbar}{4} ( F^{\alpha}_{11} + F^{\alpha}_{22} ) d^{\gamma}_{12} \frac{\partial^2 \Gamma}{\partial P^{\gamma} \partial P^{\alpha}}
	\end{eqnarray}
	The second term is
	\begin{eqnarray}
	- \frac{P^{\alpha}}{M^{\alpha}} \frac{\partial \zeta}{\partial R^{\alpha}} 
	&=& - \frac{P^{\alpha}}{M^{\alpha}} \left(
	\frac{i \hbar}{V_{11}-V_{22}}
	\frac{P^{\gamma}}{M^{\gamma}}  d^{\gamma}_{12} \frac{\partial \Delta}{\partial R^{\alpha}}
	+ \frac{i \hbar \Delta (F^{\alpha}_{11} - F^{\alpha}_{22}) }{(V_{11}-V_{22})^2}
	\frac{P^{\gamma}}{M^{\gamma}}  d^{\gamma}_{12} \right. 
	\\ \nonumber & &
	\left. + \frac{i \hbar \Delta}{V_{11}-V_{22}}
	\frac{P^{\gamma}}{M^{\gamma}}  \frac{\partial d^{\gamma}_{12}}{\partial R^{\alpha}} 
	+ \frac{i \hbar}{2} 
	\frac{\partial d^{\gamma}_{12}}{\partial R^{\alpha}} \frac{\partial \Gamma}{\partial P^{\gamma}}
	+ \frac{i \hbar}{2} 
	d^{\gamma}_{12} \frac{\partial^2 \Gamma}{\partial P^{\gamma}\partial R^{\alpha}} \right) \\
	\nonumber 
	&=& 
    - i \hbar \frac{P^{\alpha}}{M^{\alpha}} \frac{P^{\gamma}d^{\gamma}_{12}}{M^{\gamma}(V_{11}-V_{22})} \frac{\partial \Delta}{\partial R^{\alpha}}
	\\ \nonumber & &
		- i \hbar \frac{P^{\alpha}}{M^{\alpha}} \frac{P^{\gamma} d^{\gamma}_{12}}{M^{\gamma}}  \frac{ \Delta }{(V_{11}-V_{22})^2}  (F^{\alpha}_{11} - F^{\alpha}_{22})
	\\ \nonumber & &
    - i \hbar \frac{P^{\alpha}}{M^{\alpha}} \frac{\partial d^{\gamma}_{12}}{\partial R^{\alpha}} \frac{P^{\gamma}}{M^{\gamma}}   \frac{ \Delta}{V_{11}-V_{22}}
	\\ \nonumber & &
    -  \frac{i \hbar}{2}  \frac{P^{\alpha}}{M^{\alpha}}
		\frac{\partial d^{\gamma}_{12}}{\partial R^{\alpha}} \frac{\partial \Gamma}{\partial P^{\gamma}}
	\\ \nonumber & &
    - \frac{i \hbar}{2}  \frac{P^{\alpha}}{M^{\alpha}} 
		d^{\gamma}_{12} \frac{\partial^2 \Gamma}{\partial P^{\gamma}\partial R^{\alpha}} 
	\end{eqnarray}
	The third term is
	\begin{eqnarray}
	- \frac{\partial \zeta }{\partial t} 
	&=&  -i \hbar \frac{P^{\gamma}d^{\gamma}_{12}}{M^{\gamma}(V_{11}-V_{22})} \frac{\partial  \Delta}{\partial t} 
	-  \frac{i \hbar}{2} d^{\gamma}_{12} \frac{\partial}{\partial t}\frac{\partial \Gamma}{\partial P^{\gamma}} 
	\\ \nonumber 
	&=&  -i \hbar \frac{P^{\gamma}d^{\gamma}_{12}}{M^{\gamma}(V_{11}-V_{22})} \frac{\partial  \Delta}{\partial t} 
	-  \frac{i \hbar}{2} d^{\gamma}_{12} \frac{\partial}{\partial P^{\gamma}}\frac{\partial \Gamma}{\partial t} 
	\end{eqnarray}
	Now we must evaluate $\frac{\partial \Delta}{\partial t}$ and $\frac{\partial \Gamma}{\partial t}$. Applying 
	\begin{eqnarray}
	\frac{\partial A^W_{11}}{\partial t}
	& = &  \frac{2 P^{\alpha}}{M^{\alpha}} Re \left( A^W_{12} d^{\alpha}_{21} \right) - \frac{P^{\alpha}}{M^{\alpha}} \frac{\partial A^W_{11}}{\partial R^{\alpha}} - F^{\alpha}_{11}  \frac{\partial A^W_{11}}{\partial P^{\alpha}}  - Re \left( \frac{\partial A^W_{12} }{\partial P^{\alpha}}  F^{\alpha}_{21} \right) \\ \nonumber
	\frac{\partial A^W_{22}}{\partial t}
	& = &  \frac{2 P^{\alpha}}{M^{\alpha}} Re \left( A^W_{21} d^{\alpha}_{12} \right) - \frac{P^{\alpha}}{M^{\alpha}} \frac{\partial A^W_{22}}{\partial R^{\alpha}} - F^{\alpha}_{22}  \frac{\partial A^W_{22}}{\partial P^{\alpha}}  - Re \left( \frac{\partial A^W_{21} }{\partial P^{\alpha}}  F^{\alpha}_{12} \right)
	\end{eqnarray}
	we find: 
	\begin{eqnarray}
	\frac{\partial \Delta}{\partial t}
	&=& -\frac{4P^{\alpha}}{M^{\alpha}} Re\left( A^W_{12} d^{\alpha}_{21} \right) 
	- \frac{P^{\alpha}}{M^{\alpha}} \frac{\partial \Delta}{\partial R^{\alpha}} - F^{\alpha}_{22}  \frac{\partial A^W_{22}}{\partial P^{\alpha}} + F^{\alpha}_{11}  \frac{\partial A^W_{11}}{\partial P^{\alpha}} \\\nonumber
	\frac{\partial \Gamma}{\partial t}
	&=& -\frac{P^{\alpha}}{M^{\alpha}}\frac{\partial \Gamma}{\partial R^{\alpha}} - \left(F^{\alpha}_{11}  \frac{\partial A^W_{11}}{\partial P^{\alpha}} + F^{\alpha}_{22}  \frac{\partial A^W_{22}}{\partial P^{\alpha}} \right) - 2Re \left( \frac{\partial A^W_{12} }{\partial P^{\alpha}}  F^{\alpha}_{21} \right)
	\end{eqnarray}
	\begin{eqnarray}
	\frac{\partial}{\partial  P^{\gamma}}\frac{\partial \Gamma}{\partial t} 
	&=& -\frac{\partial}{\partial  P^{\gamma}}  \left(\frac{P^{\alpha}}{M^{\alpha}}\frac{\partial \Gamma}{\partial R^{\alpha}} 
	+ \left(F^{\alpha}_{22}  \frac{\partial A^W_{11}}{\partial P^{\alpha}} + F^{\alpha}_{22}  \frac{\partial A^W_{22}}{\partial P^{\alpha}} \right) 
	+ 2Re \left( \frac{\partial A^W_{12} }{\partial P^{\alpha}}  F^{\alpha}_{21} \right)\right) 
	\end{eqnarray}
	
	Thus
	\begin{eqnarray}
	- \frac{\partial \zeta }{\partial t} 
	&=&  -i \hbar \frac{P^{\gamma}d^{\gamma}_{12}}{M^{\gamma}(V_{11}-V_{22})} \frac{\partial  \Delta}{\partial t} 
	-  \frac{i \hbar}{2} d^{\gamma}_{12} \frac{\partial}{\partial P^{\gamma}}\frac{\partial \Gamma}{\partial t} 
	\\ \nonumber &=&
		i\hbar\frac{P^{\gamma}d^{\gamma}_{12}}{M^{\gamma}(V_{11}-V_{22})}
		\left(\frac{4P^{\alpha}}{M^{\alpha}} Re\left( A^W_{12} d^{\alpha}_{21} \right)  \right) 
	\\ \nonumber & &
		+  i\hbar\frac{P^{\gamma}d^{\gamma}_{12}}{M^{\gamma}(V_{11}-V_{22})} \frac{P^{\alpha}}{M^{\alpha}} \frac{\partial \Delta }{\partial R^{\alpha}} 
	\\ \nonumber & &
		+ i\hbar\frac{P^{\gamma}d^{\gamma}_{12}}{M^{\gamma}(V_{11}-V_{22})} \left(F^{\alpha}_{22}  \frac{\partial A^W_{22}}{\partial P^{\alpha}}
		- F^{\alpha}_{11}  \frac{\partial A^W_{11}}{\partial P^{\alpha}}\right) 
	\\ \nonumber & &
		+ \frac{i \hbar}{2} \frac{d^{\gamma}_{12} P^{\alpha} }{M^{\alpha}} \frac{\partial^2 \Gamma}{\partial R^{\alpha} \partial P^{\gamma}} 
	\\ \nonumber & &
		+ \frac{i \hbar}{2} \frac{d^{\gamma}_{12} }{M^{\gamma}} \frac{\partial \Gamma}{\partial R^{\gamma}} 
	\\ \nonumber & &
		+ \frac{i \hbar}{2} d^{\gamma}_{12} \left(F^{\alpha}_{11} \frac{\partial^2 A^W_{11}}{\partial P^{\alpha} \partial P^{\gamma}} 
		+ F^{\alpha}_{22} \frac{ \partial^2 A^W_{22}}{\partial P^{\alpha} \partial P^{\gamma}}  \right)
	\\ \nonumber & &
		+ i \hbar d^{\gamma}_{12}Re\left(\frac{\partial^2 A^W_{12}}{\partial P^{\alpha}\partial P^{\gamma}}F^{\alpha}_{21}\right)
	\end{eqnarray}
	Putting all three terms together, we find:
	\begin{eqnarray}
	 \frac{\partial B^W_{12} }{\partial t} 
	&=& 
	-\frac{i}{\hbar} 
	\left( V_{11} - V_{22} \right)  B^W_{12}
	- \frac{P^{\alpha}}{M^{\alpha}} \frac{\partial B^W_{12}}{\partial R^{\alpha}} 
	- \frac{1}{2} \left( F^{\alpha}_{11} + F^{\alpha}_{22} \right) \frac{\partial B^W_{12}}{\partial P^{\alpha}} 
	\\ \nonumber & &
    - \frac{i \hbar}{2} (F^{\alpha}_{11} + F^{\alpha}_{22})
	\frac{ P^{\gamma}d^{\gamma}_{12} }{M^{\gamma}(V_{11}-V_{22})} \frac{\partial \Delta }{\partial P^{\alpha} }
    \\ \nonumber & &
    - \frac{i \hbar}{2} (F^{\alpha}_{11} + F^{\alpha}_{22})
	\frac{d^{\alpha}_{12}\Delta }{M^{\alpha}(V_{11}-V_{22})}
    \\ \nonumber & &
    - \frac{i \hbar}{4} (F^{\alpha}_{11} + F^{\alpha}_{22})d^{\gamma}_{12} \frac{\partial^2 \Gamma}{\partial P^{\gamma} \partial P^{\alpha}} 
	\\ \nonumber & &
     - i \hbar \frac{P^{\alpha}}{M^{\alpha}} \frac{P^{\gamma}d^{\gamma}_{12}}{M^{\gamma}(V_{11}-V_{22})} \frac{\partial \Delta}{\partial R^{\alpha}} 
	 \\ \nonumber & &
	- i \hbar \frac{P^{\alpha}}{M^{\alpha}} \frac{P^{\gamma} d^{\gamma}_{12}}{M^{\gamma}}  \frac{ \Delta }{(V_{11}-V_{22})^2}  (F^{\alpha}_{11} - F^{\alpha}_{22})
	\\ \nonumber & &
    - i \hbar \frac{P^{\alpha}}{M^{\alpha}} \frac{\partial d^{\gamma}_{12}}{\partial R^{\alpha}} \frac{P^{\gamma}}{M^{\gamma}}   \frac{ \Delta}{V_{11}-V_{22}}
	\\ \nonumber & &
	-  \frac{i \hbar}{2}  \frac{P^{\alpha}}{M^{\alpha}}
	\frac{\partial d^{\gamma}_{12}}{\partial R^{\alpha}} \frac{\partial \Gamma}{\partial P^{\gamma}}
	\\ \nonumber & &
	- \frac{i \hbar}{2}  \frac{P^{\alpha}}{M^{\alpha}} 
	d^{\gamma}_{12} \frac{\partial^2 \Gamma}{\partial P^{\gamma}\partial R^{\alpha}}
	\\ \nonumber & &
		+i\hbar\frac{P^{\gamma}d^{\gamma}_{12}}{M^{\gamma}(V_{11}-V_{22})}
		\left(\frac{4P^{\alpha}}{M^{\alpha}} Re\left( A^W_{12} d^{\alpha}_{21} \right)  \right)
	\\ \nonumber & &
		+  i\hbar\frac{P^{\gamma}d^{\gamma}_{12}}{M^{\gamma}(V_{11}-V_{22})} \frac{P^{\alpha}}{M^{\alpha}} \frac{\partial \Delta }{\partial R^{\alpha}}
	\\ \nonumber & &
		+ i\hbar\frac{P^{\gamma}d^{\gamma}_{12}}{M^{\gamma}(V_{11}-V_{22})} \left(F^{\alpha}_{22}  \frac{\partial A^W_{22}}{\partial P^{\alpha}}
		- F^{\alpha}_{11}  \frac{\partial A^W_{11}}{\partial P^{\alpha}}\right) 
	\\ \nonumber & &
		+ \frac{i \hbar}{2} \frac{d^{\gamma}_{12} P^{\alpha} }{M^{\alpha}} \frac{\partial^2 \Gamma}{\partial R^{\alpha} \partial P^{\gamma}} 
	\\ \nonumber & &
		+ \frac{i \hbar}{2} \frac{d^{\gamma}_{12} }{M^{\gamma}} \frac{\partial \Gamma}{\partial R^{\gamma}} 
	\\ \nonumber & &
		+ \frac{i \hbar}{2} d^{\gamma}_{12} \left(F^{\alpha}_{11} \frac{\partial^2 A^W_{11}}{\partial P^{\alpha} \partial P^{\gamma}} 
		+ \frac{F^{\alpha}_{22} \partial^2 A^W_{22}}{\partial P^{\alpha} \partial P^{\gamma}}  \right)
	\\ \nonumber & &
		+ i \hbar d^{\gamma}_{12}Re\left(\frac{\partial^2 A^W_{12}}{\partial P^{\alpha}\partial P^{\gamma}}F^{\alpha}_{21}\right)
	\end{eqnarray}
	Next, we eliminate and combine a few terms to find:
	\begin{eqnarray}
	\frac{\partial B^W_{12}}{\partial t}&=& 
	-\frac{i}{\hbar} 
	\left( V_{11} - V_{22} \right)  B^W_{12}
	- \frac{P^{\alpha}}{M^{\alpha}} \frac{\partial B^W_{12}}{\partial R^{\alpha}} 
	- \frac{1}{2} \left( F^{\alpha}_{11} + F^{\alpha}_{22} \right) \frac{\partial B^W_{12}}{\partial P^{\alpha}} 
	\\ \nonumber & &
	 - \frac{i \hbar}{2} (F^{\alpha}_{11} + F^{\alpha}_{22})
		\frac{d^{\alpha}_{12}\Delta }{M^{\alpha}(V_{11}-V_{22})}
	\\ \nonumber & &
		- i \hbar \frac{P^{\alpha}}{M^{\alpha}} \frac{P^{\gamma} d^{\gamma}_{12}}{M^{\gamma}}  \frac{ \Delta }{(V_{11}-V_{22})^2}  (F^{\alpha}_{11} - F^{\alpha}_{22})
	\\ \nonumber & &
	- i \hbar \frac{P^{\alpha}}{M^{\alpha}} \frac{\partial d^{\gamma}_{12}}{\partial R^{\alpha}} \frac{P^{\gamma}}{M^{\gamma}}   \frac{ \Delta}{V_{11}-V_{22}}
	\\ \nonumber & &
		-  \frac{i \hbar}{2}  \frac{P^{\alpha}}{M^{\alpha}}
		\frac{\partial d^{\gamma}_{12}}{\partial R^{\alpha}} \frac{\partial \Gamma}{\partial P^{\gamma}}
	\\ \nonumber & &
		+i\hbar\frac{P^{\gamma}d^{\gamma}_{12}}{M^{\gamma}(V_{11}-V_{22})}
		\left(\frac{4P^{\alpha}}{M^{\alpha}} Re\left( A^W_{12} d^{\alpha}_{21} \right)  \right)
	\\ \nonumber & &
		+ \frac{i\hbar}{2}\frac{P^{\gamma}d^{\gamma}_{12}}{M^{\gamma}(V_{11}-V_{22})} \left((F^{\alpha}_{22} - F^{\alpha}_{11}) \frac{\partial \Delta}{\partial P^{\alpha}}
		\right) 
	\\ \nonumber & &
		+ \frac{i \hbar}{2} \frac{d^{\gamma}_{12} }{M^{\gamma}} \frac{\partial \Gamma}{\partial R^{\gamma}} 
	\\ \nonumber & &
		+ \frac{i \hbar}{4} d^{\gamma}_{12} \left(
		\left(F^{\alpha}_{22} - F^{\alpha}_{11}\right)
		\frac{\partial^2 \Delta}{\partial P^{\alpha} \partial P^{\gamma}}  \right)
	\\ \nonumber & &
		+ i \hbar d^{\gamma}_{12}Re\left(\frac{\partial^2 A^W_{12}}{\partial P^{\alpha}\partial P^{\gamma}}F^{\alpha}_{21}\right)
	\\ \nonumber
	\end{eqnarray}
	Lastly, we convert $A^W_{12} \rightarrow B^W_{12} + \zeta$
	\begin{eqnarray}
	\frac{\partial B^W_{12}}{\partial t}&=& 
	-\frac{i}{\hbar} 
	\left( V_{11} - V_{22} \right)  B^W_{12}
	- \frac{P^{\alpha}}{M^{\alpha}} \frac{\partial B^W_{12}}{\partial R^{\alpha}} 
	- \frac{1}{2} \left( F^{\alpha}_{11} + F^{\alpha}_{22} \right) \frac{\partial B^W_{12}}{\partial P^{\alpha}} 
	\\ \nonumber & &
	- \frac{i \hbar}{2} (F^{\alpha}_{11} + F^{\alpha}_{22})
		\frac{d^{\alpha}_{12}\Delta }{M^{\alpha}(V_{11}-V_{22})}
	\\ \nonumber & &
		- i \hbar \frac{P^{\alpha}}{M^{\alpha}} \frac{P^{\gamma} d^{\gamma}_{12}}{M^{\gamma}}  \frac{ \Delta }{(V_{11}-V_{22})^2}  (F^{\alpha}_{11} - F^{\alpha}_{22})
	\\ \nonumber & &
	 - i \hbar \frac{P^{\alpha}}{M^{\alpha}} \frac{\partial d^{\gamma}_{12}}{\partial R^{\alpha}} \frac{P^{\gamma}}{M^{\gamma}}   \frac{ \Delta}{V_{11}-V_{22}}
	\\ \nonumber & &
		-  \frac{i \hbar}{2}  \frac{P^{\alpha}}{M^{\alpha}}
		\frac{\partial d^{\gamma}_{12}}{\partial R^{\alpha}} \frac{\partial \Gamma}{\partial P^{\gamma}}
	\\ \nonumber & &
		+i\hbar\frac{P^{\gamma}d^{\gamma}_{12}}{M^{\gamma}(V_{11}-V_{22})}
		\left(\frac{4P^{\alpha}}{M^{\alpha}} Re\left( B^W_{12} d^{\alpha}_{21} \right)  \right) 
	\\ \nonumber & &
		+ \frac{i\hbar}{2}\frac{P^{\gamma}d^{\gamma}_{12}}{M^{\gamma}(V_{11}-V_{22})} \left((F^{\alpha}_{22} - F^{\alpha}_{11}) \frac{\partial \Delta}{\partial P^{\alpha}}
		\right)
	\\ \nonumber & &
		+ \frac{i \hbar}{2} \frac{d^{\gamma}_{12} }{M^{\gamma}} \frac{\partial \Gamma}{\partial R^{\gamma}} 
	\\ \nonumber & &
		+ \frac{i \hbar}{4} d^{\gamma}_{12} \left(
		\left(F^{\alpha}_{22} - F^{\alpha}_{11}\right)
		\frac{\partial^2 \Delta}{\partial P^{\alpha} \partial P^{\gamma}}  \right)
	\\ \nonumber & &
		+ i \hbar d^{\gamma}_{12}Re\left(\frac{\partial^2 B^W_{12}}{\partial P^{\alpha}\partial P^{\gamma}}F^{\alpha}_{21}\right)
	\\ \nonumber & &
	+i\hbar\frac{P^{\gamma}d^{\gamma}_{12}}{M^{\gamma}(V_{11}-V_{22})}
	\left(\frac{4P^{\alpha}}{M^{\alpha}} Re\left( \zeta d^{\alpha}_{21} \right)  \right) 
	\\ \nonumber & &
	+ i \hbar d^{\gamma}_{12}Re\left(\frac{\partial^2 \zeta}{\partial P^{\alpha}\partial P^{\gamma}}F^{\alpha}_{21}\right)
	\end{eqnarray}
	The last two terms above can be rewritten according to Eqs. \ref{eq_A11_term1} and \ref{eq_A11_term2}
	\begin{eqnarray}
	   & &
		i\hbar\frac{P^{\gamma}d^{\gamma}_{12}}{M^{\gamma}(V_{11}-V_{22})}
		\left(\frac{4P^{\alpha}}{M^{\alpha}} Re\left( \zeta d^{\alpha}_{21} \right)  \right) 
		+ i \hbar d^{\gamma}_{12}Re\left(\frac{\partial^2 \zeta}{\partial P^{\alpha}\partial P^{\gamma}}F^{\alpha}_{21}\right) 
		\\ \nonumber 
		&=& 
		i\hbar\frac{P^{\gamma}d^{\gamma}_{12}}{M^{\gamma}(V_{11}-V_{22})} \left[2\hbar Im\left(d^{\alpha}_{21}\frac{P^{\gamma}}{M^{\gamma}}d^{\gamma}_{12}\right)\frac{\partial \Gamma}{\partial P^{\alpha}}
		 \right]
		 - i\hbar d^{\gamma}_{12}\frac{\partial }{\partial P^{\gamma}} \left[\hbar Im\left( d^{\alpha}_{12} \frac{P^{\gamma}}{M^{\gamma}}  d^{\gamma}_{21}\right) \frac{\partial \Delta}{\partial P^{\alpha}} \right] 
		 \\ \nonumber 
		 &=& 
		 i\hbar\frac{P^{\gamma}d^{\gamma}_{12}}{M^{\gamma}(V_{11}-V_{22})} \left[2\hbar Im\left(d^{\alpha}_{21}\frac{P^{\gamma}}{M^{\gamma}}d^{\gamma}_{12}\right)\frac{\partial \Gamma}{\partial P^{\alpha}}
		 \right]
		 \\ \nonumber  & &
		 - i\hbar d^{\gamma}_{12} \left[\hbar Im\left( \frac{ d^{\alpha}_{12} d^{\gamma}_{21}}{M^{\gamma}} \right) \frac{\partial \Delta}{\partial P^{\alpha}} \right] 
		 \\ \nonumber  & &
		 - i\hbar d^{\gamma}_{12} \left[\hbar Im\left( d^{\alpha}_{12} \frac{P^{\gamma}}{M^{\gamma}}  d^{\gamma}_{21}\right) \frac{\partial^2 \Delta}{\partial P^{\alpha} \partial P^{\gamma}} \right] 
\end{eqnarray}
In the end, we arrive at the following expression for $\frac{\partial B^W_{12}}{\partial t}$:
\begin{eqnarray}
\frac{\partial B^W_{12}}{\partial t}&=& 
-\frac{i}{\hbar} 
\left( V_{11} - V_{22} \right)  B^W_{12}
- \frac{P^{\alpha}}{M^{\alpha}} \frac{\partial B^W_{12}}{\partial R^{\alpha}} 
- \frac{1}{2} \left( F^{\alpha}_{11} + F^{\alpha}_{22} \right) \frac{\partial B^W_{12}}{\partial P^{\alpha}} 
\\ \nonumber & &
- \frac{i \hbar}{2} (F^{\alpha}_{11} + F^{\alpha}_{22})
\frac{d^{\alpha}_{12}\Delta }{M^{\alpha}(V_{11}-V_{22})}
\\ \nonumber & &
- i \hbar \frac{P^{\alpha}}{M^{\alpha}} \frac{P^{\gamma} d^{\gamma}_{12}}{M^{\gamma}}  \frac{ \Delta }{(V_{11}-V_{22})^2}  (F^{\alpha}_{11} - F^{\alpha}_{22})
\\ \nonumber & &
- i \hbar \frac{P^{\alpha}}{M^{\alpha}} \frac{\partial d^{\gamma}_{12}}{\partial R^{\alpha}} \frac{P^{\gamma}}{M^{\gamma}}   \frac{ \Delta}{V_{11}-V_{22}}
\\ \nonumber & &
-  \frac{i \hbar}{2}  \frac{P^{\alpha}}{M^{\alpha}}
\frac{\partial d^{\gamma}_{12}}{\partial R^{\alpha}} \frac{\partial \Gamma}{\partial P^{\gamma}}
\\ \nonumber & &
+i\hbar\frac{P^{\gamma}d^{\gamma}_{12}}{M^{\gamma}(V_{11}-V_{22})}
\left(\frac{4P^{\alpha}}{M^{\alpha}} Re\left( B^W_{12} d^{\alpha}_{21} \right)  \right) 
\\ \nonumber & &
+ \frac{i\hbar}{2}\frac{P^{\gamma}d^{\gamma}_{12}}{M^{\gamma}(V_{11}-V_{22})} \left((F^{\alpha}_{22} - F^{\alpha}_{11}) \frac{\partial \Delta}{\partial P^{\alpha}}
\right)
\\ \nonumber & &
+ \frac{i \hbar}{2} \frac{d^{\gamma}_{12} }{M^{\gamma}} \frac{\partial \Gamma}{\partial R^{\gamma}} 
\\ \nonumber & &
+ \frac{i \hbar}{4} d^{\gamma}_{12} \left(
\left(F^{\alpha}_{22} - F^{\alpha}_{11}\right)
\frac{\partial^2 \Delta}{\partial P^{\alpha} \partial P^{\gamma}}  \right)
\\ \nonumber & &
+ i \hbar d^{\gamma}_{12}Re\left(\frac{\partial^2 B^W_{12}}{\partial P^{\alpha}\partial P^{\gamma}}F^{\alpha}_{21}\right)
\\ \nonumber & &
+ i\hbar\frac{P^{\gamma}d^{\gamma}_{12}}{M^{\gamma}(V_{11}-V_{22})} \left[2\hbar Im\left(d^{\alpha}_{21}\frac{P^{\gamma}}{M^{\gamma}}d^{\gamma}_{12}\right)\frac{\partial \Gamma}{\partial P^{\alpha}}
\right]
\\ \nonumber  & &
- i\hbar d^{\gamma}_{12} \left[\hbar Im\left( \frac{ d^{\alpha}_{12} d^{\gamma}_{21}}{M^{\gamma}} \right) \frac{\partial \Delta}{\partial P^{\alpha}} \right] 
\\ \nonumber  & &
- i\hbar d^{\gamma}_{12} \left[\hbar Im\left( d^{\alpha}_{12} \frac{P^{\gamma}}{M^{\gamma}}  d^{\gamma}_{21}\right) \frac{\partial^2 \Delta}{\partial P^{\alpha} \partial P^{\gamma}} \right] 
\end{eqnarray}
